\documentclass{ws-ijmpb}

\begin{document}

\markboth{Yin Zhong and Hong-Gang Luo}
{Orthogonal Dirac semimetal on honeycomb lattice and the electron spectral function in orthogonal metallic states}

%
\catchline{}{}{}{}{}
%

\title{Orthogonal Dirac semimetal on honeycomb lattice and the electron spectral function in orthogonal metallic states}

\author{Yin Zhong\footnote{Center for Interdisciplinary Studies, Lanzhou University, Lanzhou 730000, China}}

\address{Center for Interdisciplinary Studies $\&$ Key Laboratory for
Magnetism and Magnetic Materials of the MoE, Lanzhou University, Lanzhou 730000, China
\\
zhongy05@hotmail.com}

\author{Hong-Gang Luo\footnote{Center for Interdisciplinary Studies, Lanzhou University, Lanzhou 730000, China}}

\address{Center for Interdisciplinary Studies $\&$ Key Laboratory for
Magnetism and Magnetic Materials of the MoE, Lanzhou University, Lanzhou 730000, China\\
Beijing Computational Science Research Center, Beijing 100084, China\\
luohg@lzu.edu.cn}

\maketitle

\begin{history}
\received{Day Month Year}
\revised{Day Month Year}
\end{history}

\begin{abstract}
Recently, a concept of orthogonal metal has been introduced to reinterpret the disordered state of slave-spin representation in the Hubbard model as an exotic gapped metallic state. We have extended this concept to study the slave-spin representation of Hubbard model on the honeycomb lattice.[Phys. Rev. B \textbf{86}, 165134 (2012)]
It is found that a novel gapped metallic state coined orthogonal Dirac semimetal is identified. Such state corresponds to the disordered phase of slave-spin and has the same thermal-dynamical and transport properties as Dirac semimetal but its singe-particle excitation is gapped. More importantly, for generical orthogonal metallic states, the electron spectral function which may be probed in future angle-resolved photoemission spectroscopy (ARPES) experiments are studied.
\end{abstract}

\keywords{Dirac semimetal; orthogonal metal; honeycomb lattice.}

\section{Introduction} \label{intr}
Understanding the non-Fermi liquids is still a challenge for condensed matter community, which are expected to appear near the quantum criticality and should be beyond the conventional Landau Fermi liquid paradigm. \cite{Sachdev2011,Rosch,Sachdev2008}
The popular idea to attack this challenging problem is to fractionalize the electrons in the model Hamiltonian into more elementary collective excitations, e.g., quasiparticles like spinon, holon, and so on, in terms of many distinct slave-particle approaches. \cite{Wen,Senthil2003,Senthil2004,Florens,Lee2005}

Recently, an interesting quantum Monte Carlo simulation\cite{Meng} claimed a gapped spin liquid should appear in the honeycomb lattice at half-filling between the usual Dirac semimetal and antiferromagnetic insulating phases. \cite{Herbut2006,Kotov2010} Besides the elusive spin liquids which is still under hot controversy, it is also interesting to see whether a possible non-Fermi liquid state can exist in the honeycomb lattice. This issue is stimulated by a recent work of Nandkishore, Metlitski and Senthil, \cite{Nandkishore} where they reinspected the $Z_{2}$ slave-spin representation of single-band Hubbard model and pointed out that the correct disordered state of slave spins should not be a Mott insulator but an exotic fractionalized metallic state called orthogonal metal. \cite{deMedici,Hassan,Ruegg,Yu}

The orthogonal metals represent a large number of non-Fermi liquid states, which have the same thermodynamics and transport as the usual Landau Fermi liquid, but their electronic spectral function has a gap and this leads to the name 'the simplest non-Fermi liquids'. Therefore, an orthogonal metal-Fermi liquid transition instead of Mott transition should appear in the slave-spin representation of the single-band Hubbard model though no numerical studies have done on this topic. Furthermore, if the multi-orbital models like the Anderson lattice model are considered, an orbital-selective orthogonal metal transition have been found by the the present authors\cite{Zhong2012} and is argued to be an alternative Kondo breakdown mechanism for certain heavy fermion compounds. \cite{Senthil2003,Senthil2004,Paul,Pepin,Pepin2008,Paul2008}

In the present paper, we will give a brief introduction to our previous study on the $Z_{2}$ slave-spin representation of Hubbard model on the honeycomb lattice.\cite{Zhong2012b} It is found that while the ordered state of slave-spin is identified as the usual Dirac semimetal, an exotic gapped semimetal named orthogonal Dirac semimetal survives when the slave spin becomes disordered. As expected, the orthogonal Dirac semimetal has the same thermodynamics and transport as the usual Dirac semimetal. Issues about critical behaviors are not covered here and we invite readers to consult our previous paper [Phys. Rev. B \textbf{86}, 165134 (2012)] for details.

More importantly, beyond the previous studies, since it is interesting to detect the generical orthogonal metallic states in real materials, here we have studied the electron spectral function of orthogonal metallic state, which may be probed in angle-resolved photoemission spectroscopy (ARPES) experiments. The apparent distinction of the electron spectral functions from the usual Fermi liquids should provide some clues to probe such non-Fermi liquids in future experiments.

The remainder of this paper is organized as follows. In Sec. \ref{sec2}, we introduce the Hubbard model in the honeycomb lattice and reformulated it in terms of the slave spin representation. Then, a mean-field decoupling is used and two resulting mean-field states are analyzed in Sec. \ref{sec3}. One state is the expected orthogonal Dirac semimetal and the other is the usual Dirac semimetal. Thus, Sec. \ref{sec2} and \ref{sec3} completes the introduction to the orthogonal Dirac semimetal. Then, in Sec. \ref{sec4}, the electron spectral functions are studied, which may be detected in future experiments. Finally, Sec. \ref{sec5} is devoted to a conclusion which summarizes our main findings.

\section{$Z_{2}$ slave spin representation and the Hubbard model in the honeycomb lattice} \label{sec2}
The model we used is the Hubbard model in the honeycomb lattice at half-filling, \cite{Herbut2006,Ruegg,Zhong2012b}
\begin{eqnarray}
&&H=-t\sum_{\langle ij\rangle\sigma}(c_{i\sigma}^{\dag}c_{j\sigma}+h.c.)+\frac{U}{2}\sum_{i}(n_{i}-1)^{2}, \label{eq1}
\end{eqnarray}
where $n_{i}=\sum_{\sigma}c_{i\sigma}^{\dag}c_{i\sigma}$, $U$ is the onsite Coulomb energy between electrons on the same site and $t$ is the hopping energy between nearest-neighbor sites. Since we are interested in the case of half-filling, the chemical potential has been set to zero. This model has been studied by many authors and is believed to exhibit several distinct phases depending on the ratio of $U/t$. (For a review, see Ref. 11.) For small $U/t$, the usual Dirac semimetal appears with nearly free relativistic Dirac fermions being the low energy excitation. In contrast, an antiferromagnetic Mott insulator survives for large $U/t$ and spin rotation symmetry breaks spontaneously. Besides, some exotic quantum spin liquid Mott insulating states with a charge gap, e.g. algebraic spin liquid (ASL) and $Z_{2}$ spin liquid, may exist in the intermediate coupling as argued by slave-particle techniques and confirmed by sophisticated numerical simulations. \cite{Meng,Hermele2007,Wang2010,Tran2011,Lu2011,Clark}

\subsection{$Z_{2}$ slave-spin representation of the Hubbard model}
In the treatment of $Z_{2}$ slave-spin approach, the physical electron $c_{\sigma}$ is fractionalized into a new slave fermion $f_{\sigma}$ and a slave spin $\tau^{x}$ as \cite{deMedici,Ruegg}
\begin{equation}
c_{i\sigma}=f_{i\sigma}\tau_{i}^{x}\label{eq2}
\end{equation}
with a constraint $\tau_{i}^{z}+1=2(n_{i}-1)^{2}$ enforced in every site. Under this representation, the original Hamiltonian can be rewritten as
\begin{eqnarray}
&&H=-t\sum_{<ij>\sigma}(\tau_{i}^{x}\tau_{j}^{x}f_{i\sigma}^{\dag}f_{j\sigma}+h.c.)+\frac{U}{4}\sum_{i}(\tau_{i}^{z}+1)\label{eq3}
\end{eqnarray}
where $n_{i}=n_{i}^{f}=\sum_{\sigma}f_{i\sigma}^{\dag}f_{i\sigma}$. Obviously, a $Z_{2}$ local gauge symmetry is left in this representation (both slave-fermions and slave spins carrying the $Z_{2}$
gauge charge) and the corresponding low-energy effective theory should respect this. The mentioned gauge structure can be seen if $f_{i\sigma}^{(\dag)}\rightarrow \epsilon_{i}f_{i\sigma}^{(\dag)}$ and $\tau_{i}^{x}\rightarrow\epsilon_{i}\tau_{i}^{x}$ with $\epsilon_{i}=\pm1$ while the whole Hamiltonian $H$ is invariant under this $Z_{2}$ gauge transformation.

\section{Mean-field theory and orthogonal Dirac semimetal} \label{sec3}
To proceed, it is helpful to consider a mean field treatment since all fluctuations far away from critical point are gapped.

It is straightforward to derive a mean-field Hamiltonian as follows \cite{Ruegg}
\begin{eqnarray}
&&H_{f}=-\sum_{\langle ij\rangle\sigma}(\tilde{t}_{ij}f_{i\sigma}^{\dag}f_{j\sigma}+h.c.)-2\sum_{i}\lambda_{i}(n_{i}^{f}-1)^{2},\label{eq4} \\
&&H_{I}=-\sum_{\langle ij\rangle\sigma}(J_{ij}\tau_{i}^{x}\tau_{j}^{x}+h.c.)+\sum_{i}\left(\lambda_{i}+\frac{U}{4}\right)\tau_{i}^{z},\label{eq5}
\end{eqnarray}
where the Lagrange multiplier $\lambda_{i}$ has been introduced to fulfill the constraint on average,
$\tilde{t}_{ij}=t\langle \tau_{i}^{x}\tau_{j}^{x}\rangle$, $J_{ij}=t\sum_{\sigma}\langle f_{i\sigma}^{\dag}f_{j\sigma}\rangle$ and an extra self-consistent equation appears as   $\langle\tau_{i}^{z}\rangle+1=2\langle(n_{i}-1)^{2}\rangle$ due to the constraint. The decoupled Hamiltonian $H_{I}$ is an extended quantum Ising model in transverse field and $H_{f}$ describes $f$ fermions in the honeycomb lattice. Here, at the mean-field level, a further simplification can be made by setting all the Lagrange multiplier $\lambda_{i}$ to be zero, provided only non-magnetic solutions are involved and a half-filling case is considered. \cite{Ruegg} This means the constraint is not violated seriously when magnetic order is absent. Therefore, we will drop the constraint term in the mean-field Hamiltonian Eqs. (\ref{eq4}) and (\ref{eq5}) hereafter.
\subsection{Quantum Ising model in the honeycomb lattice}
Let us first focus on the quantum Ising model given by Eq.(\ref{eq5}). It is well known that the standard transverse Ising model in one spatial dimension can be exactly solved by Jordan-Wigner transformation and it has two phases with the critical exponents being the same as two-dimensional classical Ising model. \cite{Sachdev2011} Beyond one spatial dimension, to our knowledge, no exact solutions exist for the quantum Ising model until now. However, generically, one may define $\langle \tau^{x}\rangle $ as a useful order parameter and there are at least two phases in two space dimensions or beyond. One is a magnetic state with $\langle \tau^{x}\rangle \neq0$ while the other is described by a vanished $\langle \tau^{x}\rangle$ and is a disordered state with an excitation gap. Moreover, there exists a QCP between these two distinct phases, whose critical properties could be described by a quantum $\varphi^{4}$ theory,

In the case of the extended quantum Ising model in the honeycomb lattice, R\"{u}egg and Fiete \cite{Ruegg2012} found that there exist a ferromagnetic phase $\langle \tau^{x}\rangle \neq0$ and a paramagnetic phase $\langle \tau^{x}\rangle=0$ by using a 4-site cluster-mean-field approximation for the mean-field Hamiltonian. However, the paramagnetic phase named the valence-bond solid (VBS) turns out to break both lattice rotation and translation symmetry. Thus, if so, the putative quantum critical point between these two states will be unstable since the ferromagnetic phase and the valence-bond solid breaks entirely different symmetries and a first-order transition is expected generally according to the Ginzburg-Landau-Wilson paradigm.

However, we note a different slave-spin treatment, which combines with a Schwinger boson analysis instead of a cluster-mean-field approximation, favors paramagnetic phases without broken lattice rotation and translation symmetry if the onsite $U$ is not sufficiently large.\cite{Vaezi2011} Thus, we expect a paramagnetic phase without any broken symmetries in the effective quantum Ising model in the honeycomb lattice \cite{Florens,Lee2005,Nandkishore,deMedici,Hassan,Ruegg,Yu} and we may simply assume that the quantum Ising model in the honeycomb lattice has a ferromagnetic ordered phase ($\langle \tau^{x}\rangle \neq0$) and a disordered paramagnetic phase without any broken symmetries ($\langle \tau^{x}\rangle =0$) with a QCP between them.

\subsection{Hamiltonian of slave-fermion in the honeycomb lattice}
Next, we treat the mean-field Hamiltonian $H_{f}$. It is noted that $H_{f}$ is a free Hamiltonian when dropping the contribution from the constraint. Then the resulting Hamiltonian describes a Dirac semimetal which has the following formulism in the low-energy limit \cite{Kotov2010}

\begin{equation}
S_{f}=\int d\tau\int d^{2}x \sum_{\sigma}\bar{\psi}_{\sigma}\gamma_{\mu}\partial_{\mu}\psi_{\sigma}, \label{eq6}
\end{equation}
where $\gamma_{0}=I_{2}\bigotimes\sigma_{z}$,$\gamma_{1}=\sigma_{z}\bigotimes\sigma_{y}$,$\gamma_{2}=I_{2}\bigotimes\sigma_{x}$
with $I_{2}$ the $2\times2$ identity matrix, $\sigma_{x},\sigma_{y},\sigma_{z}$ being the standard Pauli matrix. The Dirac fermion $\psi_{\sigma}$ is defined as $\psi_{\sigma}=[f_{A\sigma}^{1},f_{B\sigma}^{1},f_{A\sigma}^{2},f_{B\sigma}^{2}]^{T}$ with the transpose $T$ and we also have $\bar{\psi}_{\sigma}=\psi_{\sigma}^{\dag}\gamma_{0}$. $f_{M\sigma}^{\alpha}$ ($\alpha=1,2$) is the fermion in $M=A,B$ sublattices near two nonequivalent Dirac points located at $\vec{K}=-\vec{K}'=(1,1/\sqrt{3})(2\pi/\sqrt{3})$. It is clear that this free Dirac semimetal fixed point is stable for any short-range interactions, provided the coupling parameter is not large. \cite{Herbut2006} Thus, the free Dirac semimetal fixed point with disconnected Dirac points can serve as a good starting point in the honeycomb lattice at half-filling just like the usual Fermi liquid with large Fermi surface. In addition, we should emphasize it is the slave-fermion $f_{\sigma}$ that forms a Dirac semimetal but not the physical electrons $c_{\sigma}$ since in the slave-spin approach, the physical electron is a composite particle of slave-spin and slave-fermion ($c_{\sigma}=\tau^{x}f_{\sigma}$).

Because usually it is more interesting to study the instability of Dirac semimetal of physical electrons to other states via quantum phase transitions, we here assume $f$ fermions form a Dirac semimetal whatever the phases of slave spins are, and have a sharply defined Dirac quasiparticle in the remaining parts of the present paper.

\subsection{Nature of physical electrons in the honeycomb lattice and orthogonal Dirac semimetal}
To gain some features of the physical electron in the $Z_{2}$ slave-spin representation for the honeycomb lattice, it is helpful to inspect the behavior of the quasiparticle, particularly, its single-particle Green's function or equivalently its spectral function.

In the case of $\langle\tau^{x}\rangle\neq0$ (ordered state of the corresponding quantum Ising model),
it is clear that $c_{i\sigma}\simeq\langle\tau^{x}\rangle f_{i\sigma}$, which means the slave-fermion corresponds to the physical electron when the slave-spin condensates. Therefore, the physical electron excitation is a
Dirac quasiparticle since slave-fermion forms Dirac semimetal
\begin{equation}
G(k)=Z\frac{i\gamma_{\mu}k_{\mu}}{k^{2}}\label{eq7}
\end{equation}
where the quasiparticle spectral weight is defined as $Z=\langle \tau^{x}\rangle ^{2}$ and $k^{2}=\vec{k}^{2}+\omega^{2}$ with $\omega$ being the imaginary frequency. Obviously, the obtained state with condensed slave-spins is just the usual Dirac semimetal since no fractionalized excitation will appear in physical observable. In the language of the gauge theory, the hidden $Z_{2}$ gauge field in the $Z_{2}$ slave-spin representation is confined by Higgs mechanism when the slave-spin condensates, thus only gauge singlet of fractionalized particles ($f$ or $\tau^{x}$) are allowed in physical excitations due to the confined potential generated by the $Z_{2}$ gauge field.\cite{Ruegg2012,Senthil2000}

In contrast, for a vanished $\langle \tau^{x}\rangle$ (disordered state of the slave spin) and in the low energy limit, the quantum Ising model is described by an effective $\varphi^{4}$ theory and we obtain the Green's function of slave spin as follows\cite{Nandkishore}
\begin{equation}
G_{\varphi}(k,\omega)\sim\frac{1}{\Delta^{2}+c^{2}k^{2}-\omega^{2}}.\label{eq8}
\end{equation}
The corresponding spectral function can also be easily derived with the form
\begin{equation}
A_{\varphi}(k,\omega+i0^{+})=\frac{1}{2\sqrt{\Delta^{2}+c^{2}k^{2}}}[\delta(\omega-\sqrt{\Delta^{2}+c^{2}k^{2}})-\delta(\omega+\sqrt{\Delta^{2}+c^{2}k^{2}})].\label{eq9}
\end{equation}
Then the local density of states for slave-spin behaves as
\begin{eqnarray}
N_{\varphi}(\omega)&&=\int\frac{d^{2}k}{(2\pi)^{2}}A_{\varphi}(k,\omega+i0^{+})\nonumber\\
&&=\frac{1}{2c^{2}(2\pi)^{2}}\int^{\infty}_{0}dx \frac{1}{2\sqrt{\Delta^{2}+x}}[\delta(\omega-\sqrt{\Delta^{2}+x})-\delta(\omega+\sqrt{\Delta^{2}+x})]\nonumber\\
&&=\frac{1}{2c^{2}(2\pi)^{2}}\int^{\infty}_{\Delta}dy \frac{2y}{2y}[\delta(\omega-y)-\delta(\omega+y)]\nonumber\\
&&=\frac{1}{8\pi^{2} c^{2}}\Theta(|\omega|-\Delta)\label{eq10}
\end{eqnarray}
where $\Theta(x)$ is the unit-step function. Clearly, the slave-spin acquires an excitation gap $\Delta$ and the physical electron will also acquire a gap with $Z=0$. According to the definition of the orthogonal metal in the paper of Nandkishore, Metlitski and Senthil,\cite{Nandkishore} if a state has a gap for single-particle excitation and the same thermodynamics and transport properties as Landau Fermi liquid, it could be identified as an orthogonal metal. In our case, the physical $c$ electron has an excitation gap while the $f$ fermions form Dirac semimetal. More importantly, the $f$ fermions carry both charge and spin degrees of freedom of the physical $c$ electrons, thus $f$ fermions will contribute to the thermodynamics, charge and spin transports exactly in the same way as real electrons. (The contribution of slave spins can be neglected in the low energy limit, since they are gapped in the disordered state.) Therefore, the $c$ electrons are in a gapped semimetal which is named as orthogonal Dirac semimetal.

Evidently, the orthogonal Dirac semimetal will have the same thermodynamics and transport properties as the usual Dirac semimetal, but it is noted the orthogonal Dirac semimetal is indeed a $Z_{2}$ fractionalized state with the $Z_{2}$ gauge field gapped, thus the only active actor is the slave-fermion which can be defined as a real fractionalized excitation while the usual Dirac semimetal is a confined state of $Z_{2}$ gauge field. However, unlike most of slave-particle approaches, here the fractionalization is irrelevant to the spin-charge separation and it seems this feature is general for any fractionalized states obtained in $Z_{2}$ slave-spin approach.

\section{Electron spectral functions in orthogonal metallic states}\label{sec4}
Recently, the low-energy behavior of certain spin-liquid electron spectral function is worked out and one expects such  electron spectral function with a hidden Fermi surface or Dirac points could be probed by ARPES on spin-liquid candidate materials.\cite{Tang} Since Fermi surface in the orthogonal metals or orthogonal Dirac semimetal is also hidden by disordered slave spins and it is interesting to detect such exotic metallic states in real materials, we will study the electron spectral function in generical orthogonal metallic states.

\subsection{Electron spectral function at mean-field level}
Firstly, at the mean-field level, the electron Green's function is defined as\cite{Nandkishore}
\begin{eqnarray}
G(\vec{k},\omega+i0^{+})=-\int d\vec{q}d\omega_{1}d\omega_{2}\frac{\textit{f}_{B}(\omega_{1})+\textit{f}_{F}(\omega_{2})}{\omega+i0^{+}-\omega_{2}+\omega_{1}}A_{\varphi}(\vec{q},\omega_{1})A_{f}(\vec{k}+\vec{q},\omega_{2})\label{eq11}
\end{eqnarray}
where $\textit{f}_{B}$, $\textit{f}_{F}$ are Bose and Fermi distribution functions, respectively.  $A_{f}(\vec{k},\omega)$ is the spectral function for the slave-fermion in the $Z_{2}$ slave-spin approach. For the slave-fermion with a large Fermi surface, $A_{f}(\vec{k},\omega)\approx \delta(\omega-\varepsilon_{\vec{k}})$ with $\varepsilon_{\vec{k}}$ being the energy spectrum for slave fermions.

Then, one can obtain the electron spectral function at zero temperature as
\begin{eqnarray}
A(\vec{k},\omega)=-\int d\vec{q}d\omega_{1}[\Theta(\omega+\omega_{1})-\Theta(-\omega_{1})]A_{\varphi}(\vec{q},\omega_{1})A_{f}(\vec{k}+\vec{q},\omega+\omega_{1}).\label{eq12}
\end{eqnarray}
If we are only interested in the electron spectral function at the Fermi surface, the calculation of above electron spectral function can be simplified as (assuming $\omega<0$ and considering the curvature of Fermi surface with a single-patch approximation $\varepsilon_{\vec{k}}\simeq v_{F}k_{x}+k_{y}^{2}/2m$)
\begin{eqnarray}
A(k_{F},\omega)&&\approx-\int dq_{x}dq_{y}\int^{-\omega}_{0} d\omega_{1}A_{\varphi}(q_{y},\omega_{1})\delta(\omega+\omega_{1}-v_{F}q_{x}-\frac{q_{y}^{2}}{2m_{f}})\nonumber\\
&&=-\int dq_{y}\int^{-\omega}_{0} d\omega_{1}A_{\varphi}(q_{y},\omega_{1})\nonumber\\
&&=\int^{\infty}_{0} dq_{y}\int^{-\omega}_{0} d\omega_{1}\frac{1}{\sqrt{\Delta^{2}+c^{2}q_{y}^{2}}}[\delta(\omega_{1}-\sqrt{\Delta^{2}+c^{2}q_{y}^{2}})]    \nonumber\\
&&=\int^{\infty}_{\Delta} dx\int^{-\omega}_{0} d\omega_{1}\frac{1}{\sqrt{x^{2}-\Delta^{2}/c^{2}}}\delta(\omega_{1}-x)    \nonumber\\
&&=\int^{-\omega}_{0} d\omega_{1}\frac{1}{\sqrt{\omega_{1}^{2}-\Delta^{2}/c^{2}}}\Theta(\omega_{1}-\Delta)    \nonumber\\
&&=\Theta(-\omega-\Delta)\ln(\frac{-\omega+\sqrt{\omega^{2}-\Delta^{2}}}{\Delta}).\label{eq13}
\end{eqnarray}
Therefore, the electron spectral function at the Fermi surface behaves as $\Theta(-\omega-\Delta)$, and this reflects the fact that the Fermi surface for slave-fermion is hidden by the excitation gap $\Delta$ which is shown in Fig.1. We also note that the electron spectral function is consistent with the result $A(k_{F},\omega)\propto|\omega|^{d-2}$ from the simple scaling analysis with $d$ being the spatial dimension.\cite{Nandkishore}
\begin{figure}
\includegraphics[width=0.5\columnwidth]{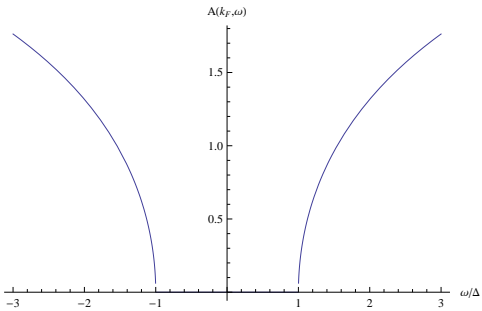}
\caption{\label{fig:1} The electron spectral function at the Fermi surface Eq. (\ref{eq13}).}
\end{figure}

Moreover, in Fig.2, we have shown a typical example for the electron spectral function in orthogonal metallic states with assuming a circular Fermi surface for the slave-fermion. The spectral function reads as follows\cite{Tang}
\begin{eqnarray}
A(\vec{k},\omega)=\sqrt{|-\omega+\Delta+\frac{(k_{x}-k_{F}\cos\phi)^{2}+(k_{y}-k_{F}\sin\phi)^{2}}{2m}|}\Theta(|\omega|-\Delta-\frac{(k_{x}-k_{F}\cos\phi)^{2}+(k_{y}-k_{F}\sin\phi)^{2}}{2m})\nonumber
\end{eqnarray}
where $\phi=\arctan(k_{y}/k_{x})$ with a $\Theta$ function denoting the existence of the gap. For comparison, the spectral function in the usual Fermi liquid state is shown in Fig.3. [Note that we only plot part of the whole momentum zone due to the $\arctan$ function.]
\begin{figure}
\includegraphics[width=0.5\columnwidth]{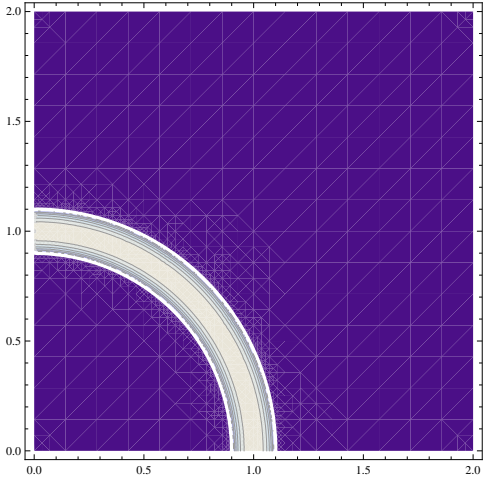}
\caption{\label{fig:1} The electron spectral function $A(\vec{k},\omega)$ with a circular Fermi surface for the slave-fermion($|\omega|=1.5>\Delta=1.49$,$k_{F}=1$ with other parameters setting to unit).}
\end{figure}

\begin{figure}
\includegraphics[width=0.5\columnwidth]{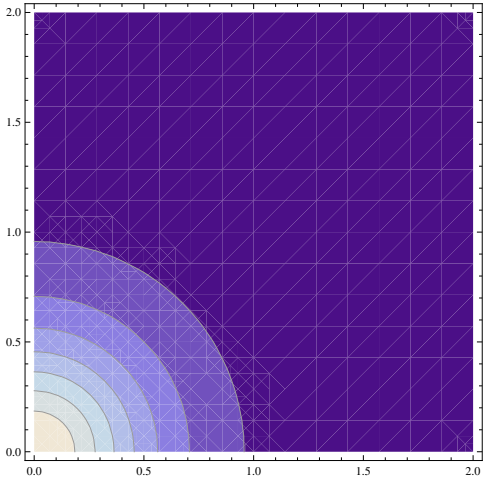}
\caption{\label{fig:1} The electron spectral function $A(\vec{k},\omega)$ in the Fermi liquid state ($|\omega|=1.5$,$k_{F}=1$ with other parameters setting to unit).}
\end{figure}

For identifying the location of Fermi surface in both the orthogonal metallic state and the usual Fermi liquid state, the electron spectral functions with $\omega=0$ are plotted in Figs.4 and 5, where no Fermi surface can be seen in the orthogonal metallic state while one observes a sharp Fermi surface in the normal Fermi liquid state.
\begin{figure}
\includegraphics[width=0.5\columnwidth]{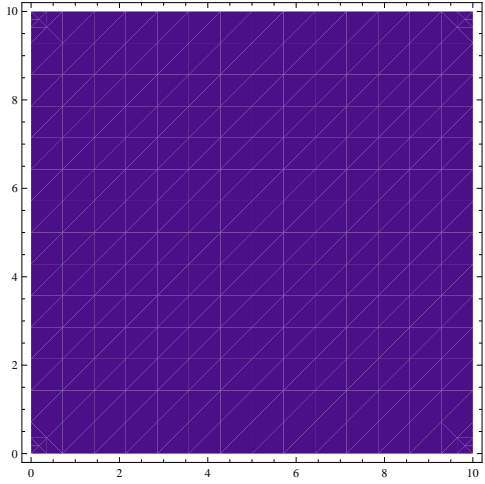}
\caption{\label{fig:1} The electron spectral function $A(\vec{k},0)$ in the orthogonal metallic state is plotted with no observable Fermi surface.}
\end{figure}

\begin{figure}
\includegraphics[width=0.5\columnwidth]{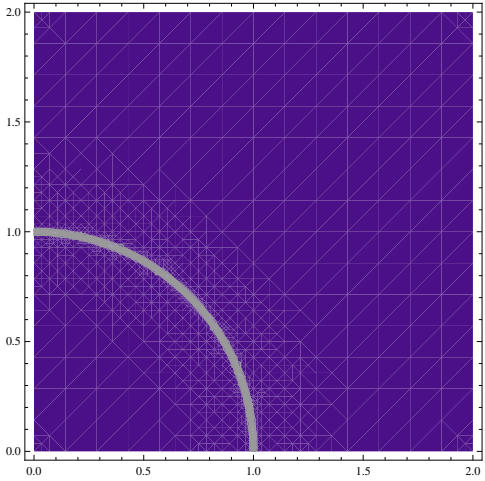}
\caption{\label{fig:1} The electron spectral function $A(\vec{k},0)$ in the Fermi liquid state is shown and a sharp Fermi surface can be seen obviously.}
\end{figure}

\section{Conclusion}\label{sec5}
In the present paper we have given a brief introduction to the orthogonal Dirac semimetal which can exist in the $Z_{2}$ slave-spin representation of Hubbard model in the honeycomb lattice at half-filling. The orthogonal Dirac semimetal survives when slave spins become disordered. This state has the same thermodynamics and transport as usual Dirac semimetal but with gapped singe-partice excitation.
More importantly, we have also studied the electron spectral function in generical orthogonal metallic states and the apparent distinction from the usual Fermi liquids should provide some clues to probe such non-Fermi liquids in future experiments.

\section*{Acknowledgements}

The work was supported partly by NSFC, the Program for NCET, the Fundamental Research Funds for the Central Universities and the national program for basic research of China.

\end{document}